\documentclass[,final]{aipproc}

\layoutstyle{6x9}

\usepackage{bm}

\newcommand{\be}{\begin{equation}}
\newcommand{\ee}{\end{equation}}
\newcommand{\bea}{\begin{eqnarray}}
\newcommand{\eea}{\end{eqnarray}}
\newcommand{\pup}{p^\uparrow}
\newcommand{\qup}{q^\uparrow}

\newcommand{\bfk}{\bm{k}}
\newcommand{\pdown}{p^\downarrow}
\newcommand{\qdown}{q^\downarrow}

\newcommand{\hup}{h^\uparrow}
\newcommand{\hdown}{h^\downarrow}

\newcommand{\NP}[1]{{\it Nucl.\ Phys.}\ {\bf #1}}

\newcommand{\PL}[1]{{\it Phys.\ Lett.}\ {\bf #1}}
\newcommand{\PR}[1]{{\it Phys.\ Rev.}\ {\bf #1}}

\begin{document}

\vspace*{-1.5cm}

\title[Hadron Spin Structure: SSA]{Hadron Spin Structure: 
Novel Effects from Transverse Single Spin Asymmetries\footnote{Talk
  delivered at the 6th Conference on 
``Quark Confinement and the Hadron Spectrum'', September
  21-25, 2004, Villasimius, Italy.} }


\vspace*{-5pt}

\author{U. D'Alesio}{
  address={Dipartimento di Fisica, Universit\`a di Cagliari, 
and INFN, Sezione di Cagliari, C.P. 170, I-09042 Monserrato (CA), Italy. 
}
}

\vspace*{-5pt}

\begin{abstract}
Transverse single spin asymmetries can be a challenging tool in
our understanding of the internal structure of hadrons. 
Some aspects and recent results are discussed.
\end{abstract}

\maketitle


\vspace*{-5pt}

Several experimental results clearly show that
transverse single spin asymmetries (SSA) in high-energy 
hadronic collisions 
can be, in particular kinematics regions, very large. 
Two relevant examples are the
transverse polarization, $P_T$, of $\Lambda$ produced in
unpolarized hadron collisions and the left-right asymmetry, $A_{N}$, observed
in $p^{\uparrow}p\to\pi\, X$:
\begin{equation} 
P_T^{\Lambda} = 
\frac{d\sigma^{AB\to \Lambda^{\uparrow} X}-
      d\sigma^{AB\to \Lambda^{\downarrow} X}}
{d\sigma^{AB\to \Lambda^{\uparrow} X}+
 d\sigma^{AB\to \Lambda^{\downarrow} X}}\quad\quad\quad
A_N=\frac{d\sigma^{A^{\uparrow}B\to C X}-
          d\sigma^{A^{\downarrow}B\to C X}}
{d\sigma^{A^{\uparrow}B\to C X} + d\sigma^{A^{\downarrow}B\to C X}}
\label{ptan}
\end{equation}
where $d\sigma$ stands for the corresponding invariant differential
cross section and $\uparrow, \downarrow$ denote transverse polarization
with respect to the hadron production plane.
While these observables can reach in size values up to 30\%-40\%, 
it is easy to see that in the usual collinear partonic kinematics,
perturbative QCD (pQCD) predicts almost vanishing SSA.
In fact at the partonic level  
single spin asymmetries are related to helicity flip amplitudes and 
to relative phases, both of which
are absent in the perturbative, chirality conserving, 
leading order interactions of quarks and gluons. 
SSA are then sensitive to higher twist 
contributions, or non perturbative effects in the long distance
physics, 
and are expected to vanish in the truly asymptotic, high-energy, large 
$Q^2$ (or $p_T$) regions.

Among the attempted explanations of $A_N$ and $P_T^\Lambda$ 
we consider here  the approach based on pQCD dynamics, through
a generalization of the factorization scheme: 
one starts from the leading twist,
collinear configuration scheme and generalizes it
with the inclusion of transverse motion of partons in distribution
functions (PDF) and hadrons in fragmentation functions (FF).
This leads, for the inclusive cross section for $AB\to CX$, to
\bea
\label{eq:kpfac1}
d\sigma^{AB \to CX} = \sum_{a,b,c,d}\hat f_{a/A}(x_a,\bfk_{\perp
  a}) \otimes \hat f_{b/B}(x_b, \bfk_{\perp b})
 \otimes\,d\hat\sigma^{ab \to cd}(\hat s, \hat t)
\otimes \hat D_{C/c}(z, \bfk_{\perp C})\,,
\eea
where $\otimes$ stands for convolution both on 
$x_i(z)$ and $\bfk_{\perp i}$. 

As discussed in \cite{dm04} the inclusion of intrinsic $\bfk_\perp$ effects 
could also play a relevant role in reducing the gap (in some cases very large) 
between the theoretical pQCD estimates and the experimental 
data for unpolarized inclusive particle production. 

For polarized processes  
the introduction of $\bfk_\perp$ and spin dependences opens up the way to
many possible spin effects; these can be summarized by the new functions:
\bea
\Delta^Nf_{q/\pup}   &\equiv& 
\hat f_{q/\pup}(x, \bfk_{\perp})-\hat f_{q/\pdown}(x, \bfk_{\perp}) 
\label{delf1} =  
\hat f_{q/\pup}(x, \bfk_{\perp})-\hat f_{q/\pup}(x, - \bfk_{\perp})  \\
\Delta^Nf_{\qup/p}  &\equiv& 
\hat f_{\qup/p}(x, \bfk_{\perp})-\hat f_{\qdown/p}(x, \bfk_{\perp}) 
\label{delf2}
 =  
\hat f_{\qup/p}(x, \bfk_{\perp})-\hat f_{\qup/p}(x, - \bfk_{\perp}) \\
\Delta^N D_{h/\qup}  &\equiv& 
\hat D_{h/\qup}(z, \bfk_{\perp}) - \hat D_{h/\qdown}(z, \bfk_{\perp}) 
\label{deld1}=   
\hat D_{h/\qup}(z, \bfk_{\perp})-\hat D_{h/\qup}(z, - \bfk_{\perp}) \\
\Delta^N D_{\hup/q}  &\equiv& 
\hat D_{\hup/q}(z, \bfk_{\perp}) - \hat D_{\hdown/q}(z, \bfk_{\perp}) 
\label{deld2} = 
\hat D_{\hup/q}(z, \bfk_{\perp})-\hat D_{\hup/q}(z, - \bfk_{\perp}) 
\>. 
\eea
The functions in Eq.s (\ref{delf1}) and 
(\ref{deld1}) are respectively the so-called Sivers \cite{siv} and
Collins  \cite{col} functions;  in Eq.s (\ref{delf2}) and (\ref{deld2})
we have  the
functions introduced by Boer and Mulders \cite{muld} and the so-called
``polarizing'' FF \cite{muld,noi3}.  
Moreover the ones in Eq.s (\ref{delf2}) and (\ref{deld1}) are
chiral-odd, while the other two are chiral-even.   
All the above functions vanish when $k_\perp=0$, are na\"{\i}vely
$T$-odd and have a clear partonic interpretation. For instance,   
the Sivers mechanism 
corresponds to the azimuthal dependence (around the
light-cone direction of the parent nucleon) of the number density of 
unpolarized partons 
inside a transversely polarized nucleon;  
the Collins mechanism 
corresponds to the azimuthal dependence (around the light-cone
direction of the fragmenting parton) of the number density of
unpolarized hadrons resulting from the fragmentation of a transversely 
polarized quark. 
Similar functions can be found in the literature 
with different notations \cite{muld,trento}.
      
In principle both the Sivers and the Collins mechanisms 
could be responsible for the
observed $A_N$ at E704 \cite{e704} (see \cite{abm}), 
while the polarizing FF in Eq.~(\ref{deld2})
could explain the measured transverse $\Lambda$
polarization in unpolarized hadron collisions \cite{noi3}.  

\begin{figure}
\hspace*{10pt}
\includegraphics[angle=-90,width = 7 cm]{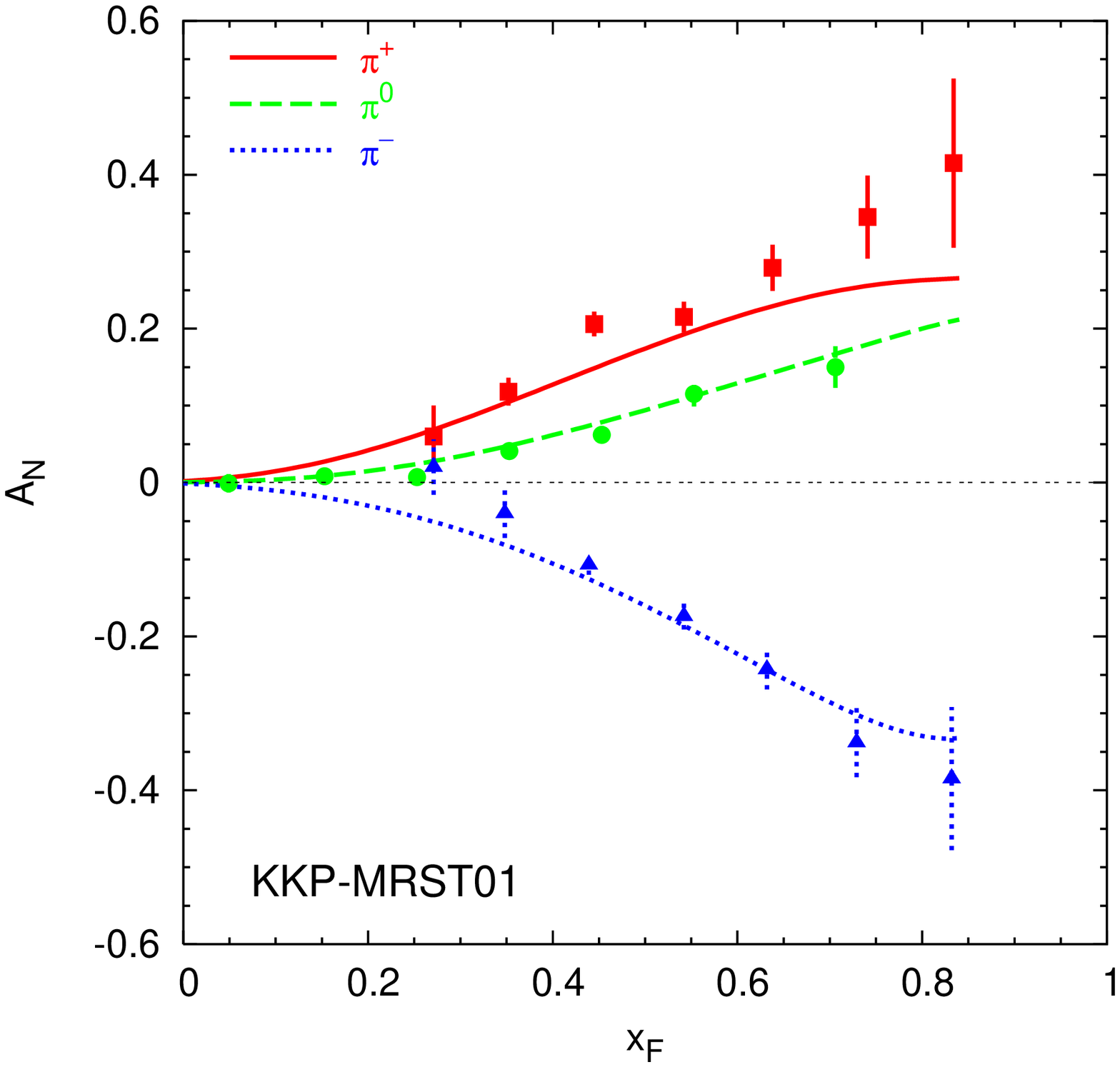}
\includegraphics[angle=-90,width = 7 cm]{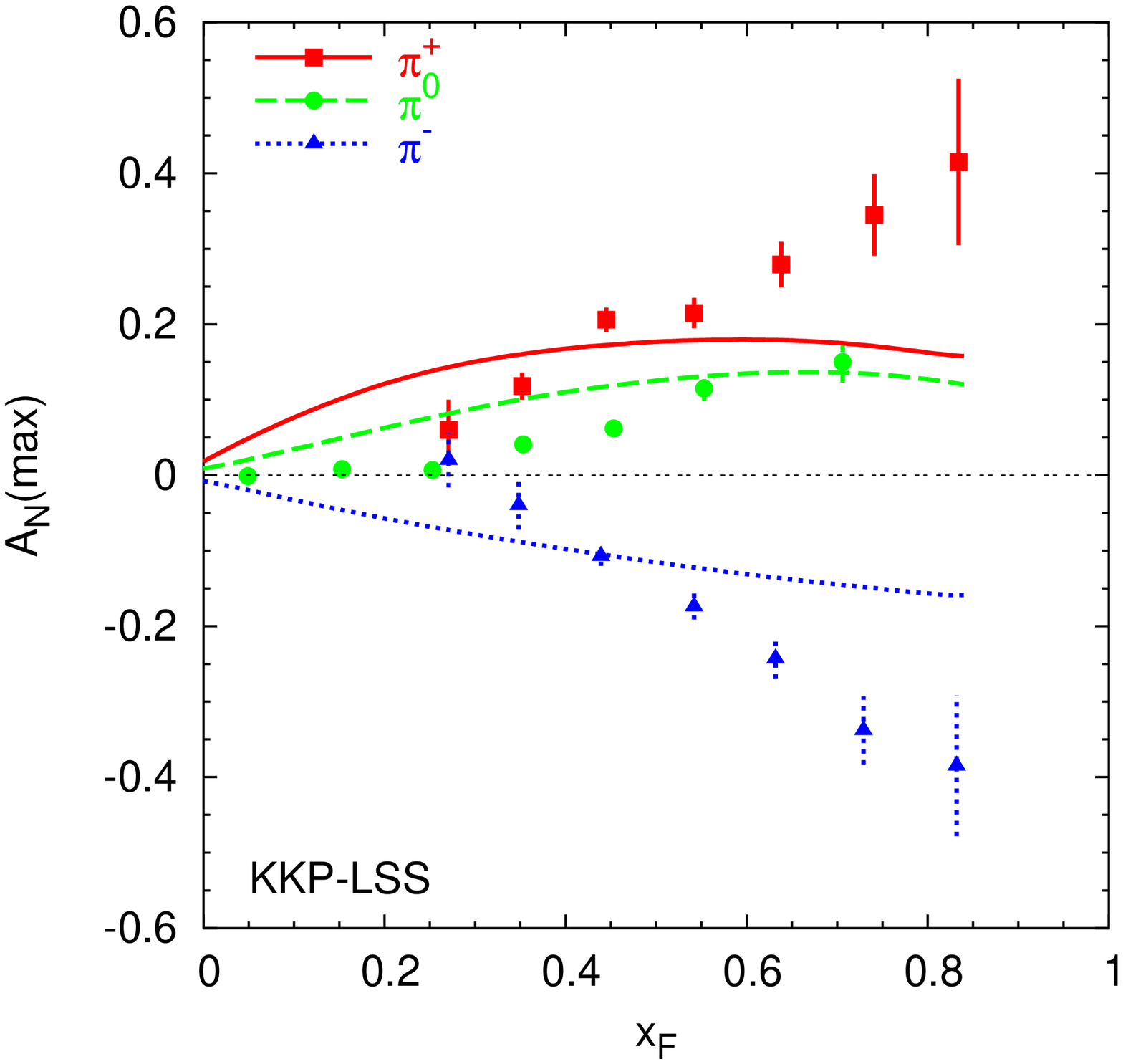}
\caption{\label{fig1}
$A_N(\pup p \to \pi X)$ vs. $x_{\!_F}$ at  $p_{\rm lab}$ = 200 GeV/$c$  
and fixed $p_{_T}$= 1.5 GeV/$c$.
Curves are obtained with Sivers effect (left plot), and  
with Collins effect (right plot). See \cite{dm04,N1} for details. 
Data are from \cite{e704}.
}
\end{figure}

Recent phenomenological studies \cite{N1}
have shown how the detailed microscopic dynamics,  
with all the correct azimuthal angular dependences, produces
a strong suppression of the transverse SSA arising from the
Collins mechanism. 
The Sivers effect is not suppressed \cite{dm04}.
In Fig.~\ref{fig1} we show our latest results for $A_N$ both in terms
of the Sivers effect alone (including only valence contributions) and with
the Collins effect alone (maximizing and including all contributions).  
A complete study of single (and double) spin asymmetries  
within the helicity formalism including $\bfk_\perp$
effects is underway \cite{noi}.

A few words on some theoretical developments are mandatory.
In fact only in the last years the role played by the gauge link
(Wilson line) entering the operator definition of these functions 
has been exploited.  
As a result we expect that 
whereas deep inelastic scattering (DIS) and Drell-Yan processes probe 
the same unpolarized PDF,
they select Sivers PDF with opposite sign \cite{newsiv}.
This poses obviously a question on universality.

As pointed out above, usually, more than one mechanism
might in principle contribute to the same SSA. Therefore it is crucial to find
proper ways to isolate each of them.

To this aim a combined experimental analysis 
of transverse SSA in Drell-Yan processes and semi-inclusive deep 
inelastic scattering (SIDIS) would be extremely
useful.    
First data on azimuthal SSA arising from Sivers effect in SIDIS are 
now available \cite{hermes} and more are
coming from HERMES and COMPASS collaborations. 
On the theoretical side it has been shown  how 
by proper suitable integration over the angular dependence of
the lepton pair, a measurement of $A_N$ in
polarized Drell-Yan processes would give a direct access 
to the (quark) Sivers function \cite{adm1}. 
A tool to learn on possible Sivers effect from gluons 
could be through the process $\pup p \to D X$ at RHIC energies 
(heavy meson production), 
being dominated by the partonic subprocess $gg \to c \bar c$ \cite{dmeson}. 

Finally,   
Collins effect could play a crucial role in the
extraction of the
the still unknown, leading twist, transversity PDF, $h_1$, 
describing the quark transverse polarization inside a transversely
polarized proton.
Indeed due to its chiral oddness, the Collins function can be a
partner of $h_1$ in the azimuthal SSA observed in 
$\ell \pup \to \ell' \pi X$. 

SSA offer a unique access to new information on
hadron structure.
This new class of spin and $\bfk_\perp$ dependent functions can give a
much deeper insight of non perturbative and long-range physics.  
New data, soon available,  will help in 
their interpretation. 
On the theoretical side, a better understanding of their fundamental
properties, like universality, QCD evolution, factorizability, 
and classification would be extremely useful.
\\

This brief overview is based on a series of papers in  
 collaboration with 
M. Anselmino, M. Boglione, E. Leader, S. Melis and F. Murgia. 

\bibliographystyle{aipproc}   

\end{document}